\documentclass[runningheads]{llncs}
\usepackage{wrapfig}
\usepackage[T1]{fontenc}
\usepackage{tabularx}
\usepackage[misc]{ifsym}
\usepackage{booktabs}
\usepackage{enumitem}
\usepackage{hyperref}
\usepackage{multirow}
\usepackage{mathtools}
\usepackage{color}
\usepackage{subcaption} 
\usepackage{xspace}
\usepackage{bm}
\usepackage{xcolor}
\usepackage{times}
\usepackage{soul}
\usepackage{url}
\usepackage[utf8]{inputenc}
\usepackage{graphicx}
\usepackage{amsmath}
\usepackage{amssymb}
\usepackage{booktabs}
\usepackage{xcolor}
\usepackage{tabularx}
\usepackage{amsmath}
\usepackage{bm}
\usepackage{subcaption}
\usepackage[ruled,linesnumbered]{algorithm2e}
\usepackage{xspace}
\usepackage{algpseudocode}
\usepackage{booktabs}
\usepackage[numbers]{natbib}
\SetKwInput{Input}{Input}
\SetKwProg{kwServer}{\textcolor{blue}{Server Update}}{}{}
\SetKwProg{kwClient}{\textcolor{blue}{Client Update}}{}{}
\algrenewcommand\alglinenumber[1]{#1}
\usepackage{graphicx}

\newcommand{\ours}{\textsc{FedCedar}\xspace}

\title{Rethinking Personalized Federated Learning with
Clustering-based Dynamic Graph Propagation}

\author{Author Name}
\author{Jiaqi Wang\inst{1}\thanks{This work was done when Jiaqi Wang interned at Visa.} \and
Yuzhong Chen\inst{2} \and
Yuhang Wu\inst{2} \and
Mahashweta Das\inst{2}\and
Hao Yang\inst{2}\and
Fenglong Ma\inst{1}}
\institute{The Pennsylvania State University \email{\{jqwang,fenglong\}@psu.edu}
\and
Visa Research \email{\{yuzchen,yuhawu,mehdas,haoyang\}@psu.edu}
}

\begin{document}
\maketitle              

\begin{abstract}

     Most existing personalized federated learning approaches are based on intricate designs, which often require complex implementation and tuning. In order to address this limitation, we propose a simple yet effective personalized federated learning framework. Specifically, during each communication round, we group clients into multiple clusters based on their model training status and data distribution on the server side. We then consider each cluster center as a node equipped with model parameters and construct a graph that connects these nodes using weighted edges. Additionally, we update the model parameters at each node by propagating information across the entire graph. Subsequently, we design a precise personalized model distribution strategy to allow clients to obtain the most suitable model from the server side. We conduct experiments on three image benchmark datasets and create synthetic structured datasets with three types of typologies. Experimental results demonstrate the effectiveness of the proposed \ours.

\keywords{Federated learning, Model personalization and distribution}
\end{abstract}

\section{Introduction}

Federated learning (FL) \cite{mcmahan2017communication} enables different data holders to cooperatively train machine learning models without sharing data. However, data heterogeneity poses a significant challenge in FL. To tackle this issue, personalized FL (PFL) has been proposed. 
Among the various research tracks in PFL, local model personalization through parameter decoupling \cite{t2020personalized} and clustering \cite{ghosh2020efficient,xie2020multi} have been explored. Parameter decoupling-based PFL has several drawbacks: (1) It demands intensive computational resources and is complex to implement. (2) It encounters the challenge of dividing the private or federated parameters \cite{arivazhagan2019federated}. In contrast, clustering-based approaches do not have such additional requirements or limitations. However, existing clustering-based approaches overlook the hidden relations between clusters and may require additional designs to handle the intricate mechanism \cite{briggs2020federated}, which complicates implementation and deployment on other frameworks. Consequently, a challenging yet practical question arises:
\emph{Is it possible to design a \textbf{simple} yet \textbf{effective} PFL framework that can address the data heterogeneity problem while considering the capture of hidden relations across clients?}

To address this question, we encounter several non-trivial challenges:
\textbf{C1: Hidden relation capturing.} In real-world applications, there may exist a physical topology among clients. Utilizing this topology effectively would contribute to local model updating. However, clients are unable to share data or access global topology information. Accurately capturing the hidden relations among clients becomes a challenging task.
\textbf{C2: Clustering-based knowledge sharing.} In FL frameworks, a random subset of clients participates in iterative model learning. Learning client-level hidden relations can be inefficient. Also, propagating under-trained models among all clients can lead to failure in FL model training. Clustering approaches allow similar clients to cooperate, which helps avoid such issues. However, most research focuses solely on gathering knowledge within clusters, disregarding information across multiple clusters. Extracting appropriate knowledge to facilitate model updates across all clusters becomes an urgent matter to address.
\textbf{C3: Fitted model acquisition.} Distributing the most fitting models back to clients as their initialization in the next communication round is not trivial. Designing an accurate model distribution strategy is crucial to maintain and inherit the benefits of iterative training, maximizing the utilization of model updates and effectively supporting model training in subsequent rounds. Moreover, the proposed approach should be easy to implement without requiring specific or complicated tuning.

To address all the aforementioned challenges, we propose a \underline{\textbf{s}}imple yet \underline{\textbf{e}}ffec\underline{\textbf{t}}ive \underline{\textbf{p}}ersonalized \underline{\textbf{fed}}erated learning framework, denoted as \ours. The framework is shown in Figure~\ref{fig:framework}.
In our approach, we first perform local model training using the respective client's data and upload the model parameters to the server. On the server side, we cluster the collected models into different groups based on their parameters. To enhance our framework, we propose a method incorporating dynamic graph construction and weighted knowledge propagation. Each cluster center is treated as a node with associated model parameters, and a graph is constructed connecting these nodes with weighted edges. Subsequently, the model parameters stored at each node are updated across the entire graph by leveraging information from other nodes through the weighted edges. Throughout the federated learning iterations, the active clients may vary, causing the clusters and graphs to change dynamically in each communication round.
Finally, we design a simple yet precise personalized model distribution strategy to maintain the benefits of iterative training and further enhance model personalization. Depending on whether an active client was selected in the previous update iteration, we assign either the personalized cluster center models or the aggregated ones to the clients.


\vspace{-2mm}
\section{Related Work}

FL is initially proposed in \cite{mcmahan2017communication}, which proposes a classical algorithm named FedAvg. Till this research, FL has recently been explored in multiple directions\cite{ma2023beyond,wang2023towards,solanki2023differentially,marchand2022securefedyj,lin2022personalized,wang2022towards,liu2022contribution,wang2023federated,liu2022fedfr,long2020federated}. Considering our proposed work, we discuss related works of model personalization and graph in FL in the following. \noindent\textbf{Personalized Federated Learning:}
Personalized FL cares more about each local model's performance, which is more sufficient and practical under the non-IID setting. One research track to solve this problem is named \texttt{FedAvg+}, which focuses on how to balance the global aggregation and local training\cite{li2020federated,fallah2020personalized,t2020personalized}. Besides the research track that we discussed, some research works focus on parameter decoupling \cite{t2020personalized} and clustering \cite{ghosh2020efficient,xie2020multi} to conduct local model personalization in FL. However, all discussed personalization FL works ignore the hidden relations between the clients or the clusters. \noindent\textbf{Graph in Federated Learning:}
Most research works, which get the graph involved in FL, focus on training GNN (Graph Neural Network) models with graph structure data distributedly\cite{wu2021fedgnn,he2021spreadgnn,zhang2021subgraph,hu2022fedgcn,lou2021stfl}. In a recent work \cite{chen2022personalized}, authors treat each local client as a node to conduct GCN learning at the server. But the number of models maintained at the server is the same as the client number, which would take intensive computation resources. Besides that, the adjacency matrix is required to be known or learned in their proposed work, which affects its practicability.

\vspace{-3mm}
\section{Methodology}

\subsection{Model Overview}
\begin{figure*}[t!]
\centering
\includegraphics[width=0.9\textwidth]{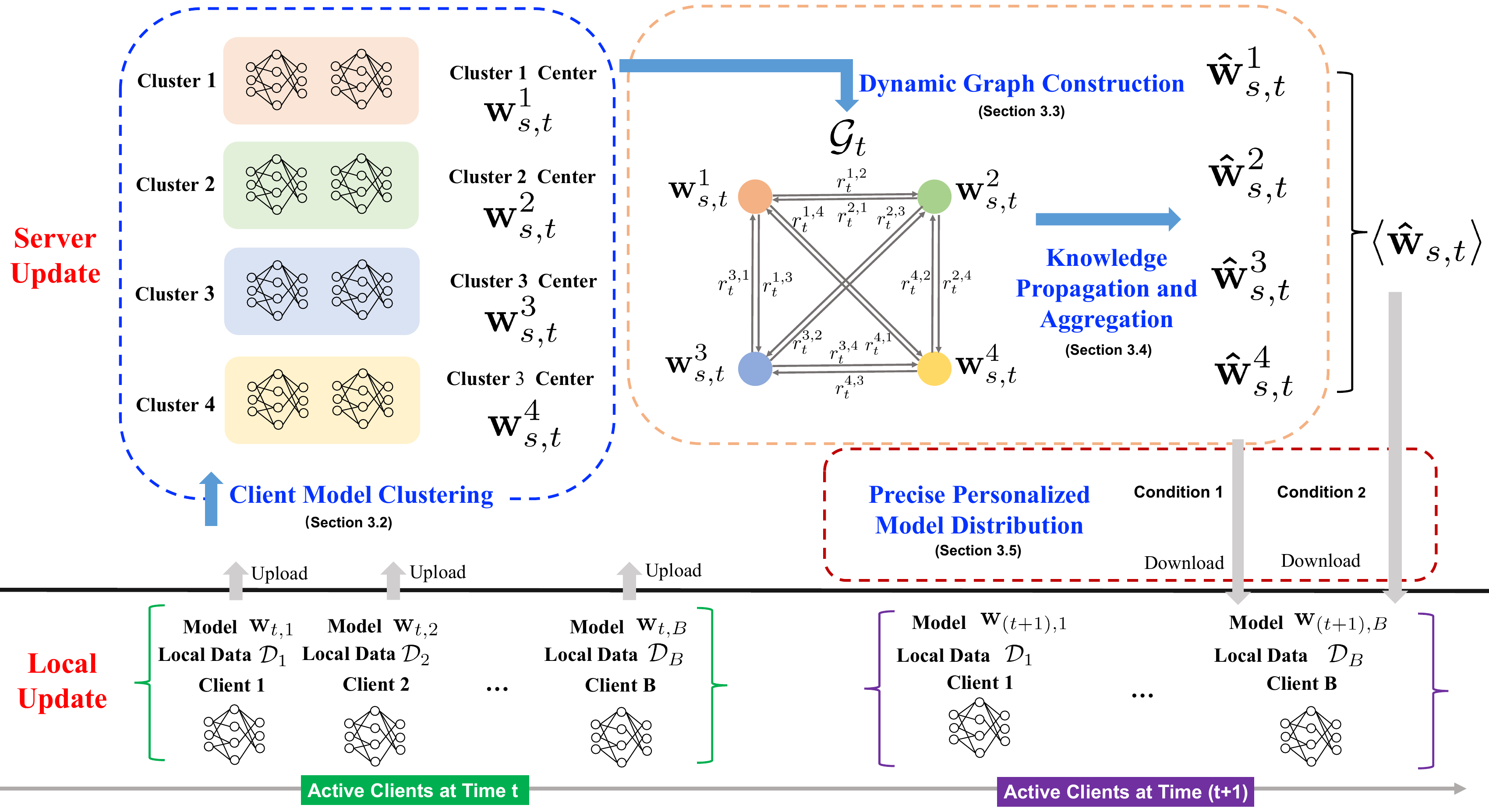}
\caption{Overview of the proposed \ours ($K = 4$ as an example). Note that clients selected at time $t$ may be different from those that are selected at time $t+1$. \textbf{Condition 1} means that the clients are selected at time $t$ as well as $(t+1)$, and \textbf{Condition 2} means that the clients are not selected at time $t$.}
\label{fig:framework}
\end{figure*}

Figure~\ref{fig:framework} shows the overview of the proposed framework, which consists of two main modules, i.e., \emph{local update} and \emph{server update}.
In the \textbf{local update} step, we first train a local model $\mathbf{w}_n$ with local data $\mathcal{D}_n$, where $\mathcal{D}_n = \{(x_i^n, y_i^n)\}_{i= 1}^{H_n}$ and $H_n$ is the number of data examples at client $n$. Specifically, we use the cross entropy (CE) loss to train $\mathbf{w}_n$ as follows: $\mathcal{L}_{n} = {\rm CE}\left(f\left(\mathbf{x}^n; \mathbf{w}_{n}\right), \mathbf{y}^n\right)$, where $f(\cdot;\cdot)$ represents the neural network, $\mathbf{x}^n$ represents the client data representations, and $\mathbf{y}^n$ is the corresponding label vector. 
Following the training procedure of general federated learning models such as FedAvg~\cite{mcmahan2017communication}, a small subset of clients with size $B$ will be randomly selected to conduct the local model learning at the $t$-th communication round. The outputs from the local client update are a set of trained model parameters $\{\mathbf{w}_{t,1},\mathbf{w}_{t,2}, \cdots, \mathbf{w}_{t,B}\}$, which will be uploaded to the server for the server update.

There are four critical components including \emph{client model clustering}, \emph{dynamic graph construction}, \emph{knowledge propagation}, and \emph{precise personalized model distribution}. After receiving $\{\mathbf{w}_{t,1}, \mathbf{w}_{t,2},\cdots, \mathbf{w}_{t,B}\}$, {\ours} will conduct the client model clustering using the K-means algorithm, i.e., dividing the $B$ uploaded clients into $K$ clusters, where $K$ is the predefined number of clusters. The outputs of this step are a set of cluster centers $\mathcal{V}_t = \{\mathbf{w}_{s,t}^2,\mathbf{w}_{s,t}^1, \cdots, \mathbf{w}_{s,t}^K\}$, where $\mathbf{w}_{s,t}^k$ denotes the center of the $k$-th cluster at the $t$-th communication round. In the dynamic graph construction step, {\ours} treats each cluster center as a node and constructs a dynamic weighted graph $\mathcal{G}_t=(\mathcal{V}_t,\mathcal{E}_t, \mathcal{R}_t)$, where $\mathcal{E}_t \subseteq \mathcal{V}_t \times \mathcal{V}_t$ is the set of edges between nodes, and $\mathcal{R}_t$ is the set of weights on edges. {\ours} then executes the knowledge propagation step across the whole graph $\mathcal{G}_t$ to update node representations, i.e., model personalization learning. Finally, {\ours} distributes the learned personalized models $\{\mathbf{\hat{w}}_{s,1}^{1}, \mathbf{\hat{w}}_{s,1}^{2}, \cdots, \mathbf{\hat{w}}_{s,t}^{K}\}$ back to the corresponding clients with the precise model distribution strategy. 
Next, we describe the details of the four components in the server update step.

\vspace{-2mm}

\subsection{Client Model Clustering}
\label{sec:model_cluster}

At the $t$-th communication round, the $B$ selected clients will upload their models $\{\mathbf{w}_{t,1}, \cdots, \mathbf{w}_{t, B}\}$ to the server. These models will be grouped into $K$ clusters using K-means \cite{long2022multi}\cite{ghosh2020efficient} by optimizing the following loss function:
\begin{equation}
\label{eq:kmeans}
    J_t = \sum_{k=1}^{K}\sum_{b=1}^{B} \delta_{t,b}^k(||\mathbf{w}_{s,t}^{k} - \mathbf{w}_{t, b}||^{2}),
\end{equation}
where $||\mathbf{w}_{s,t}^{k} - \mathbf{w}_{t,b}||^{2}$ is the Euclidean distance between the $k$-th center $\mathbf{w}_{s,t}^{k}$ and the client model $\mathbf{w}_{t,b}$. $\delta_{t,b}^k$ is the indicator. If the $b$-th client belongs to the $k$-th cluster, then $\delta_{t,b}^k =1$. Otherwise, $\delta_{t,b}^k =0$.
Note that for each communication round $t$, we will generate a new set of $K$ cluster centers $\{\mathbf{w}_{s,t}^1, \cdots, \mathbf{w}_{s,t}^K\}$.

\vspace{-2mm}
\subsection{Dynamic Weighted Graph Construction}
\label{sec:dynamic_weighted}

The clustering process helps local clients with similar data distributions or characteristics to aggregate together based on their model parameters. However, simply utilizing the results of the clustering algorithm to conduct the personalization in FL \cite{xie2020multi,ghosh2020efficient} has several limitations. 
On the one hand, the cluster centers $\{\mathbf{w}_{s,t}^1, \cdots, \mathbf{w}_{s,t}^K\}$ are obtained by averaging the client parameters within each cluster. In other words, the personalization information is only from the clients within each cluster by modeling the \emph{inner-cluster} characteristics, which ignores that \emph{between clusters}.
On the other hand, the performance of existing clustering approaches is mainly determined by the quality of the cluster assignment. One low-quality cluster may lead to slow convergence and poor overall performance. When modeling the relations between clusters, high-quality information can flow to other clusters, which may neutralize the negative effect of low-quality clusters. Thus, it is essential to model the hidden relations between clusters.

Towards this end, we build a dynamic weighted graph to capture the hidden relations between the local models to enhance the mode updates further. However, it is highly non-trivial to design a novel mechanism satisfying the following conditions and requirements. First, in the FL setting, it would be better not to increase the network transmission load due to the constraints of communication cost. Second, in a typical FL framework, we randomly sample clients at each communication round, which means different batches of clients contribute to the updates. Then the center model $\mathbf{w}_{s,t}^k$ in each cluster carries different information with huge differences. How to leverage the iterative heterogeneous knowledge appropriately is a challenge. FedAvg-based aggregation approach~\cite{mcmahan2017communication} enables rough information sharing by obtaining one global model, which cannot handle the data or model heterogeneity. 
Thus, solving such challenges requires a more reliable and convincing aggregation strategy to maintain the personalization property for each local client.

Specifically, we use $\mathcal{G}_t=(\mathcal{V}_t,\mathcal{E}_t, \mathcal{R}_t)$ to denote a graph, where $\mathcal{V}_t = \{\mathbf{w}_{s,t}^1, \cdots, \mathbf{w}_{s,t}^K\}$ is the node set, and each node corresponds to a cluster. $\mathcal{E}_t \subseteq \mathcal{V}_t \times \mathcal{V}_t$ is the set of edges, where any pair of nodes exists an edge with a corresponding weight. $\mathcal{R}_t$ is the set of edge weights. The weight is defined as the normalized cosine similarity between two cluster centers $\mathbf{w}_{s,t}^i$ and $\mathbf{w}_{s,t}^j$, where $\sum_{j=1}^{K} r^{i,j}_t = 1$.

\vspace{-5mm}
\subsection{Knowledge Propagation and Aggregation}
\label{sec:knowledge_progagation}
\vspace{-2mm}
After we build the graph with the weighted edges, we are able to reveal and describe the hidden relations between cluster centers corresponding to a set of model parameters. The next question is how we could utilize the connection to help the model update across the graph appropriately. To answer this question, we design an effective approach to enable the model parameters saved at each node to gather information from other nodes for knowledge propagation across the whole graph. 

Mathematically, given a constructed graph $\mathcal{G}_t=(\mathcal{V}_t,\mathcal{E}_t, \mathcal{R}_t)$, for $\forall k \in K$, we update the center in a weighted sum way as follows:
\begin{equation}\label{update}
\begin{split}
    \mathbf{\hat{{w}}}_{s,t}^{k} &= g([\mathbf{w}_{s,t}^{1,P-1}, \cdots, \mathbf{w}_{s,t}^{K,P-1}], [r_t^{k, 1}, \cdots, r_t^{k, K}], P) 
    = \sum_{i=1}^K r_t^{k,i}\mathbf{w}_{s,t}^{i, P-1}\\
    &= \sum_{i=1}^K r_t^{k,i} \sum_{j=1}^K r_t^{i,j}\mathbf{w}_{s,t}^{j, P-2}
    = \cdots
    = \underbrace{\sum_{i=1}^K r_t^{k,i}\sum_{j=1}^K r_t^{i,j} \cdots \sum_{z =1}^K r_t^{u,z}\mathbf{w}_{s,t}^{z, 0}}_{k \leftarrow i \leftarrow j \leftarrow \cdots \leftarrow u \leftarrow z (\text{\# path = $P$})},
\end{split}
\end{equation}
where $g$ is the function to conduct the knowledge propagation with the updated model parameters, and $P$ denotes the times we repeat the propagation process. $\mathbf{w}_{s,t}^{i, P-1}$ is the updated model parameter of node $i$ after $P-1$ rounds of propagation and aggregation, $r_{k,i}$ is the weight between node $k$ and $i$. 
When $P=1$, $\mathbf{w}_{s,t}^{z,0} = \mathbf{w}_{s,t}^{z}$, which is the output from Section 3.3. 
After this procedure, we have a set of new models named $\{\mathbf{\hat{w}}_{s,t}^{1}, \mathbf{\hat{w}}_{s,t}^{2},\cdots,\mathbf{\hat{w}}_{s,t}^{K} \}$ with the weighted information from in-cluster clients and out-of-cluster nodes.

\subsection{Precise Personalized Model Distribution}
\label{sec:precise_personalized}
After we get $K$ customized models at the server side, it is challenging to select and distribute the most fitted model to each individual active client to achieve the optimal local personalization performance at the next communication round. There are two reasons:  (1) To simulate the real-world scenario, we randomly sample clients at each communication round with the sample ratio $\gamma$. In this case, each sampled client set could vary a lot. (2) We still expect to maintain the model generalization and personalization with the benefit of the iterative training mechanism in FL. To solve the above challenges, after obtaining $\{\mathbf{\hat{w}}_{s,t}^{1}, \mathbf{\hat{w}}_{s,t}^{2},\cdots,\mathbf{\hat{w}}_{s,t}^{K} \}$, we design a precise personalized model distribution strategy.

Specifically, we need to consider two conditions with respect to the client sampling at two consecutive communication rounds $t-1$ and $t$. Correspondingly, there are sampled client set named $\mathcal{C}_{t-1}$ and $\mathcal{C}_{t}$ at time $t-1$ and $t$, respectively. 
Given $\mathcal{C}_{t-1}$ and $\mathcal{C}_{t}$, for client $n$ at communication round $t$, different model distribution strategies are implemented under the following two conditions:
\begin{itemize}
    \item \textbf{Condition 1}: if $n \in \mathcal{C}^* = \mathcal{C}_{t-1} \cap \mathcal{C}_t$ (\emph{intersection set}) at time $t$,  we trace back to the cluster $k$, to which client $n$ belongs at time $t-1$. Then we distribute model $\mathbf{\hat{w}}_{s,t-1}^{k}$ to client $n$ as the model initialization at time $t$;
    \item \textbf{Condition 2}: if $n \in \mathcal{C}^\prime = \mathcal{C}_{t} - \mathcal{C}_{t-1}$ (\emph{difference set}) at time $t-1$, we distribute 
$\langle\mathbf{\hat{w}}_{s,t-1}\rangle$ to client $n$ as the initial model parameters, where $\langle\mathbf{\hat{w}}_{s,t-1}\rangle$ is calcualted as follows: $\langle\mathbf{\hat{w}}_{s,t-1}\rangle = \frac{1}{K}(\mathbf{\hat{w}}_{s,t-1}^{1}+\mathbf{\hat{w}}_{s,t-1}^{2}+\cdots+ \mathbf{\hat{w}}_{s,t-1}^{K})$.
\end{itemize}



\section{Experiment}
\subsection{Experiment Setup}

\subsubsection{Dataset Preparation}
We conduct experiments for the image classification task on \textbf{MNIST}, \textbf{SVHN}, and \textbf{CIFAR-10} datasets in both IID and non-IID data distribution settings, respectively. 
We split the training datasets into $80\%$ for training and $20\%$ for testing. For the IID setting, the training and testing datasets are both randomly sampled. For the non-IID setting, we divide the training dataset following the approach used in \cite{chen2022personalized} and set the \textit{shard} = 2, which is an extreme non-IID setting. To test the personalization effectiveness, we sample the testing dataset following the label distribution as the training dataset. 
\vspace{-3mm}
\subsubsection{Baselines}

The proposed {\ours} is a personalized federated learning algorithm. To achieve personalization, we adopt the clustering technique and construct dynamically weighted graphs. To fairly evaluate the proposed {\ours}, we use the following baselines: \emph{(1) Classical FL Models}:
\textbf{FedAvg} \cite{mcmahan2017communication} and
\textbf{FedProx} \cite{li2020federated}; \emph{ (2) Personalized FL Models}: 
\textbf{pFedMe} \cite{t2020personalized} and
\textbf{pFedBayes} \cite{zhang2022personalized}; \emph{(3) Graph-based FL Model}: 
\textbf{SFL} \cite{chen2022personalized}; \emph{(4) Clustering-based FL Models}:
\textbf{IFCA} \cite{ghosh2020efficient} and 
 \textbf{FedSem}\cite{xie2020multi}.





\begin{table*}[t]
\centering
\caption{Performance comparison with baselines.}
\resizebox{0.85\textwidth}{!}{
\begin{tabular}{l|l|cccccc} 
\toprule 

\multirow{2}{*}{\textbf{Category}}&\textbf{Dataset}& \multicolumn{2}{c}{\textbf{MNIST}}&
\multicolumn{2}{c}{\textbf{SVHN}} & \multicolumn{2}{c}{\textbf{CIFAR-10}}\\

&\textbf{Setting}    & IID & non-IID& IID & non-IID & IID & non-IID\\
\midrule
 \multirow{2}{*}{Classical}&FedAvg & 96.77\%  & 91.02\% & 82.65\% & 81.04\% & 68.93\% & 59.85\% \\
 &FedProx & 97.87\% & 93.98\%& 83.09\% & 82.68\% & 71.07\% & 64.07\% \\
\midrule
 \multirow{3}{*}{Personalization}&Per-FedAvg
              & 96.57\% &\underline{93.56\%} & 87.74\%& 87.09\%  & 71.32\% & 75.56\%\\
 &pFedMe  & 96.90\% & 93.10\%& 87.79\% & 86.37\% & 73.92\% & 77.21\%\\
 &pFedBayes & 97.23\% &92.08\%&\underline{90.69\%} & \underline{88.03\%} & \underline{80.45\%} & \underline{79.05\%}\\
\midrule
Graph&SFL  &  96.88\% & 93.10\%& 86.97\% & 86.15\% & 79.92\% & 75.44\%\\
\midrule
\multirow{2}{*}{Cluster}&IFCA  &  \textbf{97.96\%}  & 92.09\%  & 84.44\% & 83.26\% & 79.21\% & 73.08\% \\

 &FedSem &  96.52\%  & 93.05\%  &84.75\%& 84.96\% & 79.87\% & 71.22\% \\
\midrule
Ours &\ours & \underline{97.90\%}  & \textbf{95.96\%} &  \textbf{91.19\%}& \textbf{88.76\%} & \textbf{82.80\%} & \textbf{81.53\%}\\
\bottomrule 
\end{tabular}
}
\label{tab:result}
\end{table*}
\vspace{-3mm}
\subsubsection{Implementation Details}
In our experiments, we leverage convolutional neural networks (CNNs) as our basic models for the three image datasets. For the MNIST and SVHN datasets, the network structure consists of three convolutional layers and two fully-connected layers. For the CIFAR-10 dataset, we have six convolutional layers and two fully-connected layers. All the baselines and \ours use the same networks to keep the comparison fair. Following \cite{wang2023knowledge}, the total client number is $N=100$, and the active client ratio in each communication round is 30\%. The total communication round between the server and local clients is $T =100$. The local training batch size is 16, the local training epoch is 5, and the local training learning rate is 0.01. The number of knowledge propagation $P$ is set to 2. The cluster number $K$ is set to 5. We provide the hyperparameter study to explore how the key hyperparameters affect the performance of the proposed \ours in the subsection \ref{subsec:hyper}.
\vspace{-3mm}
\subsection{Performance Evaluation}
\vspace{-2mm}
 We run each approach three times and report the \textit{average accuracy} in Table \ref{tab:result}. There are several observations and discussions as below: (1) Overall, our proposed \ours outperforms baselines on MNIST, SVHN, and CIFAR-10 datasets under both IID and non-IID settings except for the result of IFCA on MNIST under the IID setting. This is due to the fact that the MNIST dataset under the IID setting is a relatively easy task where all approaches achieve comparable performance, compared with other datasets and settings; (2)  If we focus on SVHN and CIFAR-10, we find that the advantage of our algorithm is becoming dominant, with the dataset being more complicated under the non-IID setting. Compared with FedAvg under the non-IID setting, our approach increases the accuracy rate by 9.53\% and 36.22\%, respectively. Compared with the best performance of other algorithms on SVHN (88.03\%) and CIFAR-10 (79.05\%) under the non-IID setting, our algorithm boosts the performance to 88.76\% ($\uparrow 0.73\%$) and 81.53\%($\uparrow 2.48\%$), respectively; (3) Among all baselines, pFedBayes performs the best on both SVHN and CIFAR-10 datasets. pFedBayes is specifically tailored for personalization. However, it requires the exchange of data distribution information between the clients and the server, which raises concerns about communication costs and privacy leakage.

\subsection{Ablation Study}
In this subsection, we conduct the ablation study to investigate the contribution of key modules in our proposed model \ours. We use the following reduced or modified models as baselines: \textbf{AS-1}: \emph{Without clustering}. We treat each client as a node to build the graph and conduct the knowledge propagation. Compared with \cite{chen2022personalized}, we do not have the complicated GCN learning process or the optimal global model but distribute its own model via our precise model distribution strategy. \textbf{AS-2}: \emph{Without building a graph or knowledge propagation}. We only conduct client clustering, and other modules are kept. Different from \cite{ghosh2020efficient}, we only conduct one clustering process rather than an iterative optimization to estimate the cluster identities. \textbf{AS-3}: \emph{Wthout precise personalized model distribution}. After conducting dynamic graph construction and knowledge propagation, we aggregate the models from each node into one single model and distribute it back to the clients.

\begin{wraptable}{r}{0.5\textwidth}
\vspace{-0.15in}
\caption{Ablation study. }
\resizebox{0.5\textwidth}{!}{
\begin{tabular}{l|cccccc} 

\toprule 
\textbf{Dataset}&  \multicolumn{2}{c}{\textbf{MNIST}}  & \multicolumn{2}{c}{\textbf{SVHN}} &  
             \multicolumn{2}{c}{\textbf{CIFAR-10}} \\
\hline
\textbf{Setting}            & IID & non-IID & IID & non-IID& IID & non-IID\\
\midrule
AS-1 &96.79\% &92.16\%&82.66\%&82.11\%&70.10\%&72.96\%\\
AS-2 &96.80\%& 92.53\%&82.71\%&81.39\%&70.45\%&69.15\%\\
AS-3 &97.04\%&92.01\%&83.66\%&81.05\%&71.08\%&65.37\%\\
\midrule
\ours           & \textbf{97.90\%}  & \textbf{95.96\%} &  \textbf{91.19\%}& \textbf{88.76\%} & \textbf{82.80\%} & \textbf{81.53\%}\\
\bottomrule 
\end{tabular}
}
\label{tab:ablation}
\end{wraptable}



Table \ref{tab:ablation} shows the experimental results, where we can observe that reducing one or more key modules in \ours leads to performance degradation. There are several observations and discussions as below: (1) The results compared with AS-1 show that treating each client as a node in the graph may not be able to extract enough supporting knowledge from other clients. It also shows that the help of similar clients within one cluster can enhance the model performance; (2) In AS-2, we construct the clusters and maintain the precise personalized distribution strategy. The reduction of the performance can tell the effectiveness of utilizing the hidden relations between each cluster via graph construction and knowledge propagation. It reveals that the extra weighted knowledge is able to lift the model performance; (3)The performance of AS-3 is the worst compared with AS-1 and AS-2. If we directly aggregate all the models at the nodes into one global model via doing an average, it essentially can be simplified as FedAvg.

\vspace{-3mm}
\subsection{Case Study}
To further explore the effectiveness of the proposed \ours, we design a case study to demonstrate the process of the graph formulation with respect to the communication round. We are interested in testing if our model is able to capture and reconstruct the hidden relations between local clients through iterative updates in \ours framework.
\vspace{-4mm}
\subsubsection{Structured Data Construction}
\begin{figure}[ht!]
\centering
\includegraphics[width=0.9\textwidth]{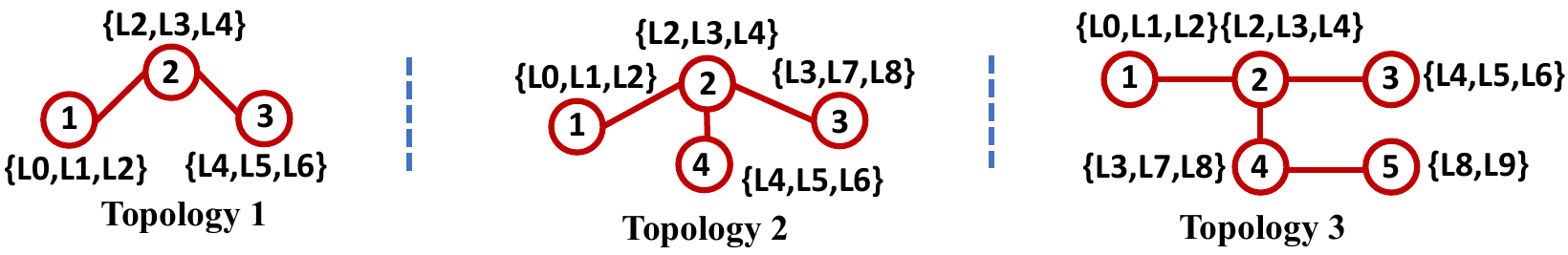}
\caption{Structured datasets with synthetic topologies.}
\label{fig:topology}
\end{figure}

To achieve our target, we structure a synthetic dataset via MNIST by building known relations between the nodes shown in Figure \ref{fig:topology}, and the goal is to test whether our graph construction method has the capability of identifying these pre-designed relations. For each node in the graph, we have 20 clients equipped with designed label distributions.  If there are common labels among a pair of nodes, an edge connects them. For example, in Topology 1, label 2 (denoted as L2) is the common label between node 1 and node 2, so there is an edge connecting the two nodes. In this case study, we construct three topologies with 3 nodes, 4 nodes, and 5 nodes, respectively. 

\vspace{-10mm}
\subsubsection{Periodic Activation Strategy}
For each node in the graph, we have 20 clients, where each client has the corresponding labels from the node as we discussed. Given a periodic round set $\mathcal{Q}$, if $t \in \mathcal{Q}$, we sample the clients from each node at the activation ratio $\gamma$ to formulate the active client set, which guarantees the formulated client set covers the label information from all the structured nodes in the graph.
At other communication rounds $t \not \in \mathcal{Q}$, we randomly sample $B$ clients from $N$ clients without considering where the clients come from, where $B = N * \gamma$. Thus, the total number of active clients is defined in this simulation experiment as follows: $\text{if $t \in \mathcal{Q}$: }\mathcal{C}_t = 
    \sum_{i = 1}^{K}(S_i * \gamma)$; $\text{if $t \not \in \mathcal{Q}$: }\mathcal{C}_t = B$, where $S_i$ is the client set at node $i$. Later, we track the graph construction performance at the round set $\mathcal{Q}$ through the communication rounds compared with the synthetic topology structure.


\vspace{-5mm}
\subsubsection{Evaluation Metrics}
We leverage the random index \cite{rand1971objective} as the evaluation metric to represent the performance of the constructed clusters across the iterative update process. It measures the similarity of the two clustering outcomes by considering all pairs of samples: it counts pairs that are assigned in the same or different clusters in the generated and ground-truth clusters \cite{hubert1985comparing}. Specifically in our case, given a set $\{\mathbf{\hat{w}}_{s,t}^{1},\mathbf{\hat{w}}_{s,t}^{2},\cdots, \mathbf{\hat{w}}_{s,t}^{K}\}$, assuming we have two partitions $\mathcal{A} = \{\mathit{A}_1,\mathit{A}_2,...,\mathit{A}_a\}$ and $\mathcal{B} = \{\mathit{B}_1,\mathit{B}_2,...,\mathit{B}_b
\}$, where there are $a$ and $b$ subsets in each partition respectively, we have: $\mathbf{R} = (\alpha + \beta) / \binom{K}{2}= 2 \times (\alpha + \beta) / K(K-1)$, where $\alpha$ and $\beta$ are the number of
agreements between the two partitions that the pairs belong to the same subsets or different subsets. Intuitively, it tells the occurrence or the probability that the partitions agree on the data sample pairs. Thus, the score $\mathbf{R} = 1$ means the perfect clustering, and $\mathbf{R} = 0$ means poor clustering.
\vspace{-3mm}
\subsubsection{Results Analysis}
\begin{wrapfigure}{r}{0.4\textwidth}
\vspace{-0.15in}
  \centering
  \includegraphics[width=0.4\textwidth]{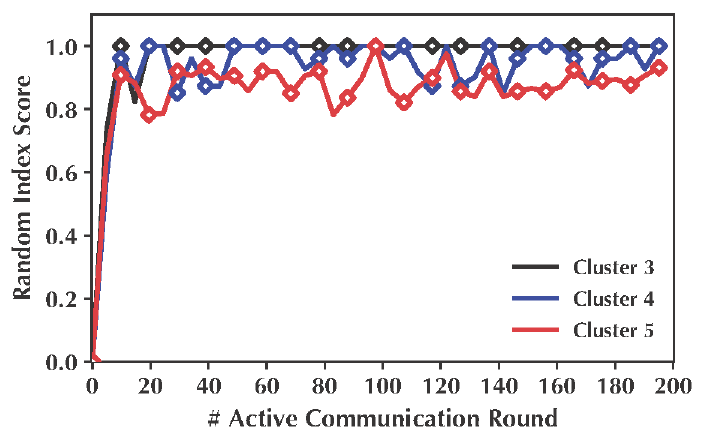}
  \caption{Random index.}
  \vspace{-0.15in}
  \label{fig:random_index}
\end{wrapfigure}
We run the case study with different cluster numbers $K = 3, 4$, and $5$. We choose to activate the targeted clients from the designed topology every 5 communication rounds, which means we have 40 data points during the total 200 communication rounds ($\mathcal{Q} = \{5,10,\cdots,200\}$). To have a better visualization, we plot the data with the scatter every 2 data points, which means we totally have 20 markers along with each line.

The results are shown in Figure \ref{fig:random_index}. In the figure, the black, blue, and red demonstrate the results with the number of clusters equal to 3, 4, and 5, respectively. When $K=3$, which is Topology 1, we can observe that there is very limited fluctuation at the first initial communication rounds. The lowest score is 0.73 and the score converges to 1 after several updates. For $K=4$, the convergence speed is slower than the setting when $K=3$, but it is close to 1 finally. Besides that, when $K=5$, its convergence speed and score are not as good as when $K=3$ or $4$. There are several possible reasons: (1) the complexity of the topology will increase the difficulty of generating the correct clusters and graphs; (2) with more nodes in our setting, there are more labels and clients involved in the updates, which increases the heterogeneity of the datasets. Generally, we observe that the random index score under three settings all reach convergence and converge to a high value, demonstrating the effectiveness and interpretability of our proposed algorithm.

\vspace{-2mm}
\subsection{Hyperparameter Study}
\label{subsec:hyper}
In this subsection, we investigate how the hyperparameters $P$ and $K$ affect the performance of \ours. $P$ controls how many times we conduct knowledge propagation across the graph at each communication round in Eq.~\eqref{update}. $K$ controls the number of clusters when we group the clients by optimizing Eq.~\eqref{eq:kmeans}. To explore the parameter sensitivity, we alter $P$ and $K$ as $\{1,2,3,4,5\}$ and $\{3,4,5,6,7\}$, respectively. 

\begin{wrapfigure}{r}{0.35\textwidth}
\vspace{-0.15in}
  \centering
  \includegraphics[width=0.35\textwidth]{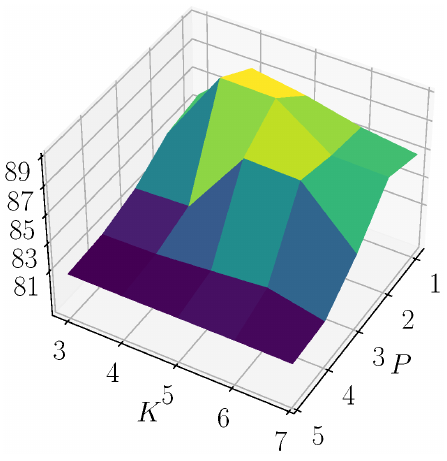}
  \caption{Hyperparameter study.}
  \vspace{-0.15in}
  \label{fig:hyper}
\end{wrapfigure}

We report the three-time average experiment results in Figure \ref{fig:hyper}. For the experiment results on SVHN, we observe: (1) The highest accuracy is $88.83\%$ appearing at $\{P = 2, K =4 \}$. (2) In general, with the increase of $P$, the accuracy shows non-monotonous behavior by firstly going up and then going down. For example, given $K = 4,5,6,$ or $7$, the best performance appears at the setting with $P = 2$ or $3$. (3) With the cluster number $K$ becomes larger, appropriate increase of $P$ is able to boost the performance. Specifically, when the cluster number $K$ increase from 3 to 6, the best performance appears at the point when $P = 1,2$ and $3$, respectively. 


\vspace{-3mm}
\section{Conclusion}

We design a personalized federated learning framework \ours equipped with the functionalities of clustering, dynamic weighted graph learning, and precise personalized model distribution to solve the data heterogeneity challenge. In particular, we build dynamic weighted graphs to conduct model aggregation and enable clients to obatain models precisely with our designed strategy. Our proposed \ours outperforms state-of-the-art baselines on three datasets for the image classification task. To our best knowledge, this is the first FL research work to embed the clustering and dynamic weighted graph construction to explore the hidden relations between clients and thus obtain the optimal personalized local models via precise personalized model distribution strategies. In future works, we would like to provide theoretical convergence and generality analysis of our model under more flexible real-world settings.

\bibliographystyle{splncs04}
\bibliography{mybibliography}

%




\end{document}